\documentclass[3p,twocolumn]{elsarticle}
\usepackage{epstopdf}
\usepackage{amssymb}
\usepackage{subfigure}
\journal{Nuclear Instruments and Methods}

\cortext[cor1]{Corresponding author.\\
E-mail address:hanna@physics.mcgill.ca (D. Hanna) }
\begin{document}
\begin{frontmatter}
\title{A Directional Gamma-Ray Detector Based on Scintillator Plates}
\author{D. Hanna\corref{cor1}, L. Sagni\`eres, P. Boyle, and A. MacLeod}
\address{Department of Physics,
McGill University,
Montreal, QC H3A 2T8, Canada} 
 
\begin{abstract}
A simple device for determining the azimthal location of a source of
gamma radiation, using ideas from astrophysical gamma-ray burst detection, is 
described. A compact and robust detector built from eight identical modules, 
each comprising a plate of CsI(Tl) scintillator coupled to a photomultiplier 
tube, can locate a point source of gamma rays with degree-scale precision 
by comparing the count rates in the different modules. 
Sensitivity to uniform environmental background is minimal.
\end{abstract}
\end{frontmatter}

\section{Introduction}
\label{sec:intro}
During the development of a Compton imager for safety and security 
applications~\cite{laurel,pat,audrey,laurel-2} we became interested in the 
idea of using a simple
detector to establish an approximate ($\pm$ 5 degrees) direction 
to the source of gamma 
radiation that was to be imaged. 
(The Compton imager had a field of view that extended $\pm$ 45 degrees in azimuth 
and $\pm$ 20 degrees in the vertical direction.)  
The imager could then be set up such that its field of view would be centred on 
this direction, thus improving the quality of initial images and saving time 
in making precise measurements of the size and location of the source.

The challenge of determining the direction to a gamma-ray source using
a minimal amount of data has already been met in the astrophysical
study of gamma-ray bursts (GRBs), which are fast (durations of a few
seconds) and non-repeating events where a flux of gamma rays, all from
the same direction, impacts a space-borne detector.  The Burst and
Transient Source Experiment (BATSE) detector~\cite{batse} was an
instrument flown on the Compton Gamma-ray Observatory
(1991-2000) and did much to improve our understanding of this
phenomenon.  BATSE comprised a set of eight large-area NaI(Tl)
scintillation counters installed on the corners of the spacecraft.
The scintillators had areas that were large with respect to their
thickness and were oriented such that they each covered an octant of
the total solid angle. The direction to a GRB was estimated by
comparing count rates in the different counters. The idea was that a
counter facing the GRB would present a larger area to it and therefore
have a higher count rate than would a counter facing in a direction
perpendicular to the GRB. Angular resolution of approximately 1$\circ$
was obtained with this method with the exact number flux
dependent.

We have used this idea to design a detector that can measure the
direction to a source of gamma rays.  The Octagonal Directional
Detector (ODD) consists of eight plates of scintillator arranged as an
eight-sided cylinder.  As such, the detector is mechanically easier to
build and deploy than one employing the same geometry as BATSE, but it
is restricted to making measurements in a single plane.  

We note that there are other ways to build a directional gamma-ray 
detector, for example with two or more different scintillators viewed by 
a single photomultiplier tube (PMT)~\cite{shirakawa}. 
Such detectors rely on careful analysis of pulse-height spectra and are
therefore more complicated than the device described here.

Due to its inherent simplicity, most features of the detector can be easily 
simulated or calculated analytically; however, we decided to build an example
instrument to check for any surprises. 
In the following we present design and construction details, followed by 
results from laboratory tests.

\section{Instrument Design}
\label{sec:design}
\begin{figure}[ht]
\centerline{\includegraphics[width=0.9\linewidth]{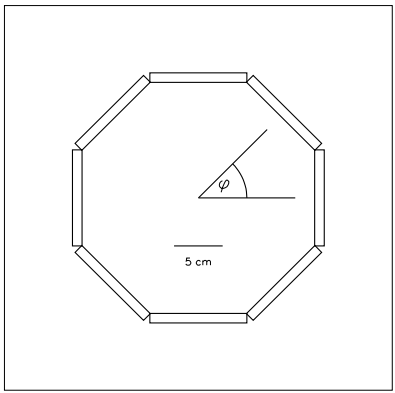}}
\vspace*{-2.0cm}
\caption{Schematic of the detector (plan view) indicating the relative
positions of the 10 cm by 20 cm by 1 cm plates of CsI(Tl) scintillator 
and the convention used for defining the azimuthal angle $\varphi$.} 
\label{plan_view}
\end{figure}

Our detector is built from
eight plates of CsI(Tl) scintillator\cite{proteus}, 
each 10~cm wide by 20~cm high and 1~cm thick.
A plan-view schematic is shown in Figure~\ref{plan_view}.
The plates are supported by a thin (1.5 mm)
aluminum structure, interior to the detector, which defines the
octagonal geometry. Light from the scintillators is detected by PC30CW5
PMTs from Sens-Tech Sensor
Technologies\cite{sens-tech}.  These devices are each equipped
with a Cockcroft-Walton high-voltage (HV) generator controlled by a
potentiometer, obviating the need for an external HV supply and
associated cables. The scintillator-PMT combination is referred to here as
a module.

\begin{figure}[ht]
\centerline{\includegraphics[width=0.9\linewidth]{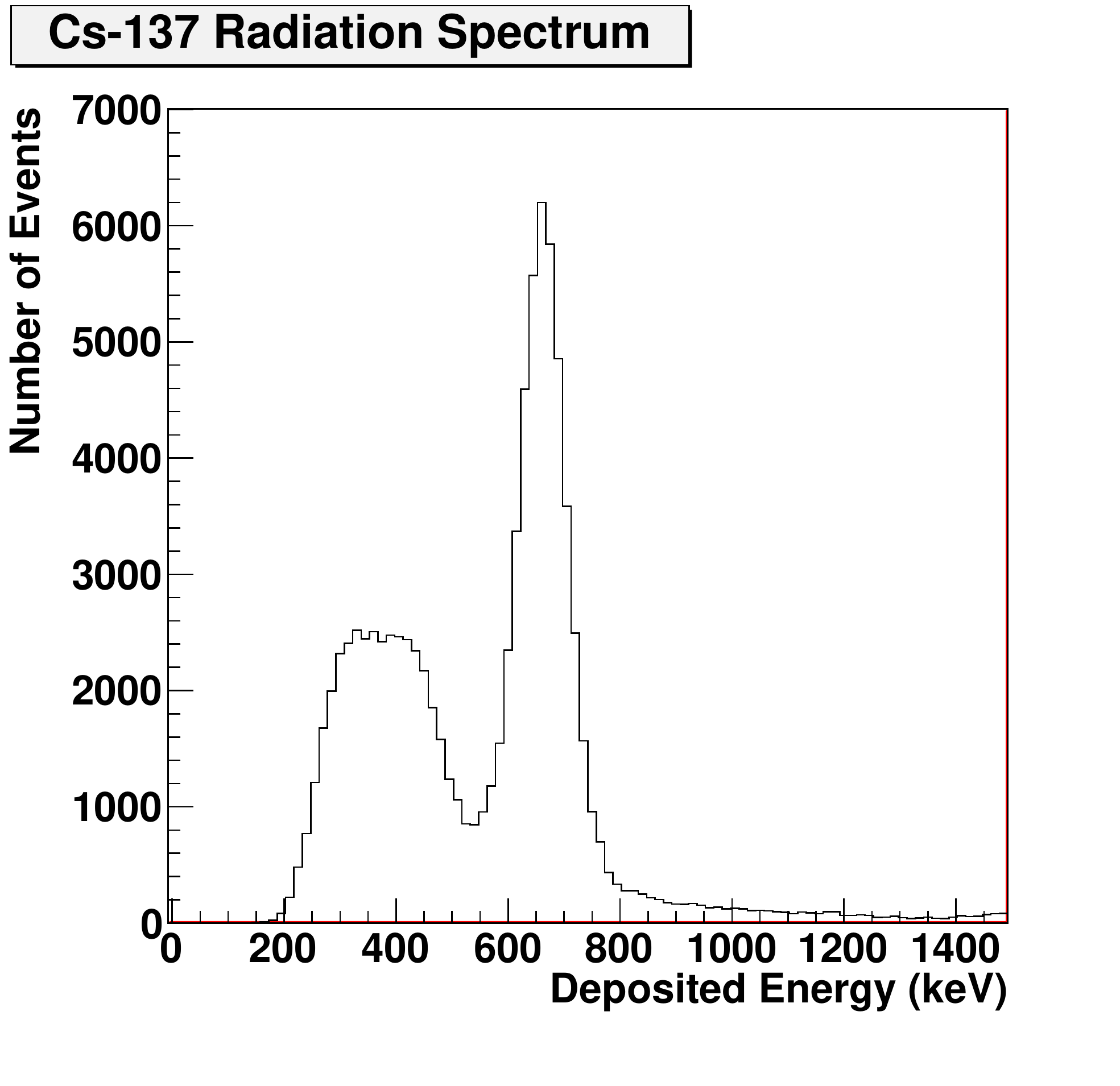}}
\vspace*{0.0cm}
\caption{Response spectrum for a single module exposed to a 
Cs-137 (662 keV) gamma-ray source. }
\label{spectrum}
\end{figure}

CsI(Tl) was chosen for its good stopping power, high light output, and
moderate cost.  Also, unlike NaI(Tl), it is only mildly 
hygroscopic.  A negative feature, for our purposes, is its long
decay time ($\tau = 1~\mu$s).  This necessitated the use of
custom pulse-shaping electronics, where the PMT pulse is shortened
using an integrator/differentiator combination.
The circuit produces two identical outputs.
One is discriminated using a comparator circuit and a NIM
logic level is produced if a user-defined threshold is crossed.  The
other is left in analog form to be used in studies involving
charge levels.

Charge information is useful for monitoring the gains of the different
channels but is not strictly necessary for the functioning of the
detector which is fundamentally a counting instrument.  A typical
charge spectrum for a module is shown in
Figure~\ref{spectrum}. 

The NIM-level pulses are counted using a
custom-built rate meter, originally designed to study fast
astrophysical optical transients~\cite{trendy}. 
This device was chosen since it was available and very portable.
Conventional NIM scalers or any fast counting electronics can also be used.

\section{Laboratory Tests}

Laboratory tests were made using a 10 $\mu$Ci Cs-137 source located
at a distance of 80 cm from the centre of the detector and at a height of 10 cm
from the base of the detector.
The flux at the detector was approximately 4.6 photons per cm$^2$.

The detector was mounted on a turntable to allow it to be rotated and this 
enabled us to perform a scan in azimuth.
Before data taking began,
a 1 $\mu$Ci Cs-137 source was placed at the centre of the detector and
the HV values for the PMTs were adjusted to equalize the count rates of the
modules.
Sixteen sets of data were acquired with the angle between the source and the 
detector changed by $\pi/8$ between each set. 
Each data set comprised 300 sets of nine scaler values,
one value for each detector module and one value representing the 
sum of counts
from a 1 MHz pulse generator.
The counting interval for each set was nominally two seconds but there were
small differences from set to set.
Data from the 1 MHz pulser were used to correct for this and 
the module counts were then converted to rates.
The rates varied between 100 Hz and 200 Hz depending on the orientation of the scintillator plate
with respect to the source.

\subsection{Detector Response Curves}

\begin{figure}[ht]
\centerline{\includegraphics[width=1.1\linewidth]{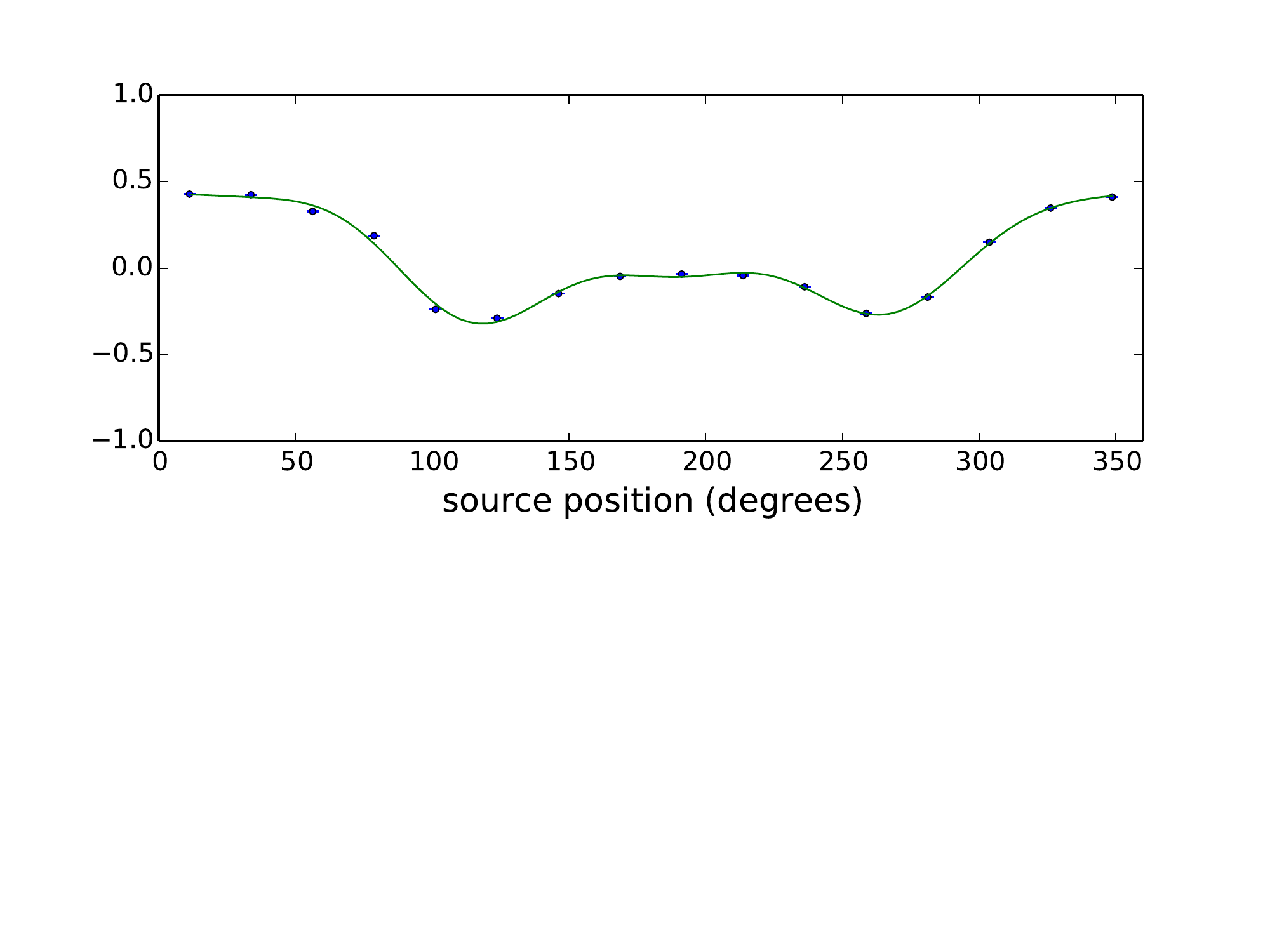}}
\vspace*{-3.0cm}
\caption{Detector response for a single module for 16 azimuthally equidistant 
source positions.
To emphasize the modulation, 
count rates have been divided by the average value and 1.0 has been subtracted.
The fit is an eleven-component Fourier series.} 
\label{module-1}
\end{figure}

The first fifty data sets for each angle were combined to improve statistical
precision and the rates plotted vs. angle setting for each counter.
A sample curve is shown in Figure~\ref{module-1}.
The count rates have been divided by their mean value and 1.0 has been 
subtracted.
This eliminates a common offset, leaving the modulation component that we
exploit for direction finding.
An eleven-component Fourier series has been fit to the data. 

The response curve reaches its global maximum when the source is directly in
front of the module and its two minima when the source is approximately 90 degrees away 
from this position.
A local maximum is obtained when the source is directly behind the module.
Here the solid angle is maximal but the module on the opposite side
of the detector absorbs some of the radiation, thus reducing the count rate.

\begin{figure}[ht]
\centerline{\includegraphics[width=1.2\linewidth]{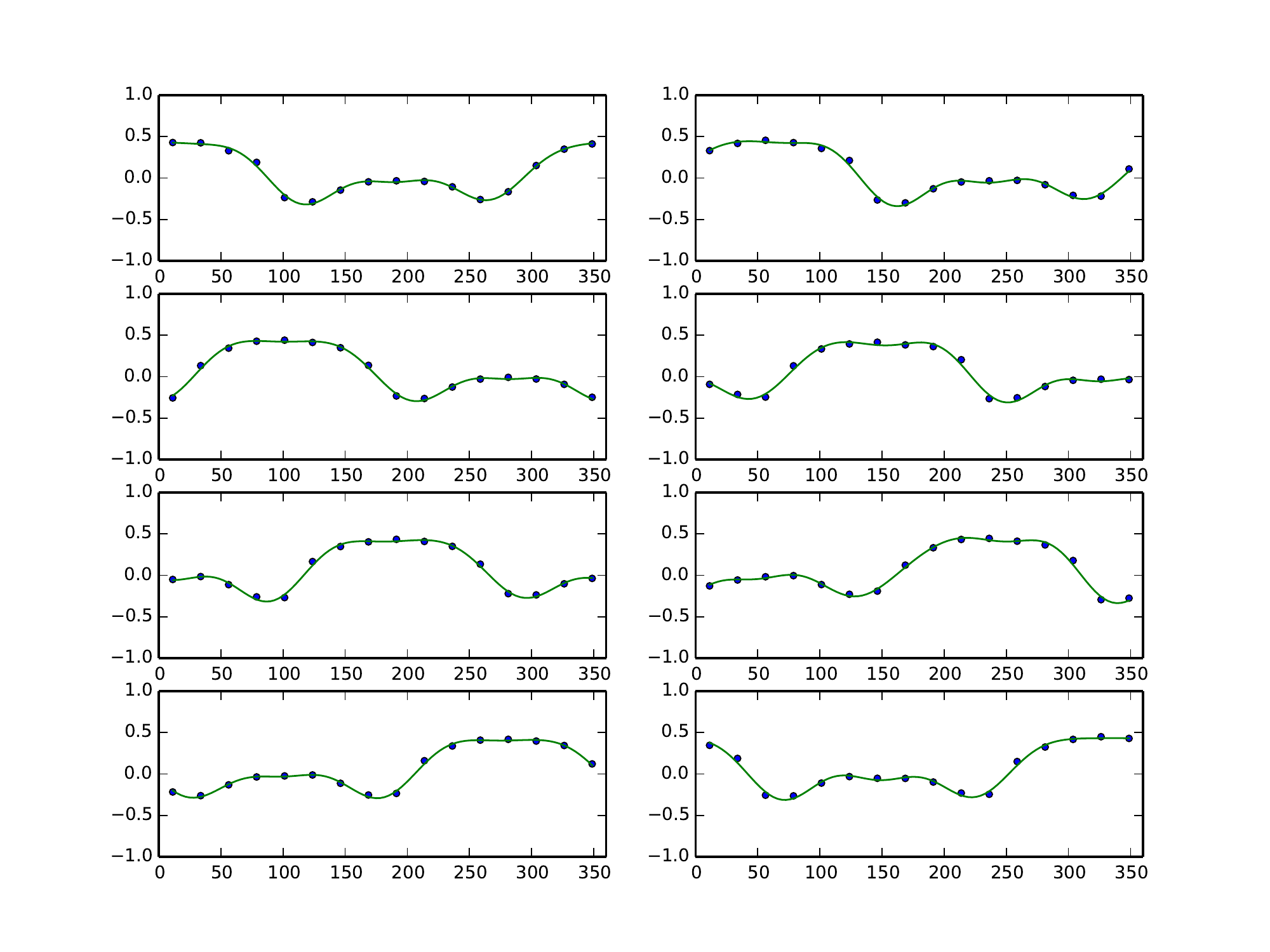}}
\vspace*{0.0cm}
\caption{As in Figure~\ref{module-1} but for all eight modules.}
\label{modules-1-8}
\end{figure}

The response curves for all eight modules are shown in Figure~\ref{modules-1-8}.
The curves are very similar, differing primarily by a phase which is
the result of the relative orientation of the module.

\begin{figure}[ht]
\centerline{\includegraphics[width=0.9\linewidth]{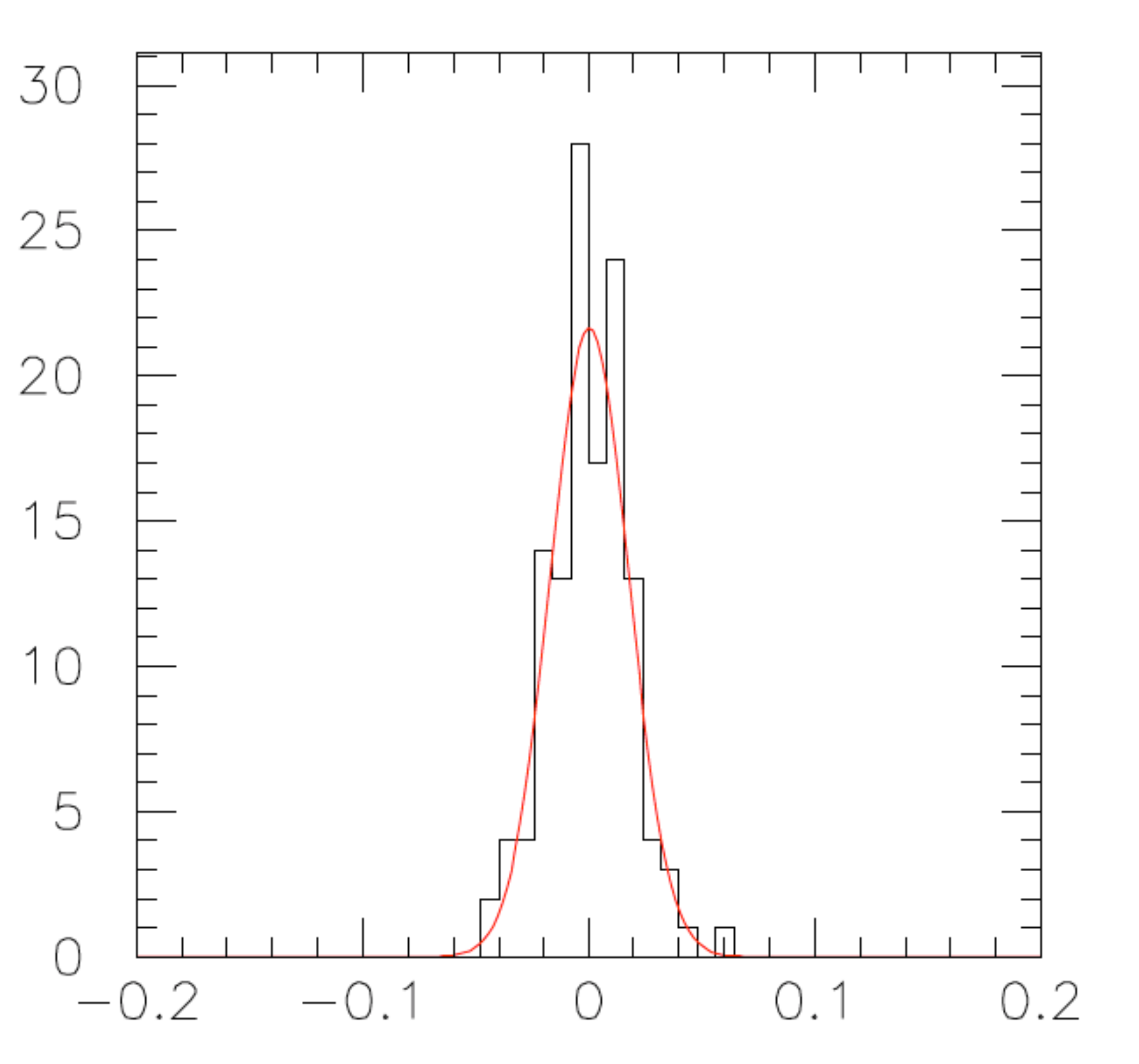}}
\vspace*{0.0cm}
\caption{A histogram of the deviations of angle-shifted calibration points from their averages,
as described in the text, demonstrating the similarity of the calibration 
curves shown in Figure~\ref{modules-1-8}. The horizontal scale is the same as the
vertical scales in Figures~\ref{module-1} and ~\ref{modules-1-8}.  
The fitted Gaussian has $\sigma = 0.017 \pm 0.002$.}
\label{self-sim}
\end{figure}

We can quantify this by shifting the 16 calibration points for each module by 
the appropriate multiple of 45 degrees, computing the average value
of the eight points at each angle, and plotting the residuals from these
averages, as in Figure~\ref{self-sim}. 
The resulting distribution is well fit by a Gaussian with 
$\sigma = 0.017 \pm 0.002$, which is consistent with the average value 
of the uncertainities on the calibration points.

\subsection{Single-Source Detection}

Having used the first 50 sets of count-rates vs. source position
to produce the calibration curves shown in Figure~\ref{modules-1-8}, 
we used the remaining  250 sets, in various combinations, 
to study single-source detection.

To determine the angle of a source from the rates one
minimizes the $\chi^{2}$, which is given by \\

$\chi^{2} = \Sigma (r_{i}-f_{i}(\phi))^{2}/\sigma_{i}^{2}$. \\

\noindent 
Where $r_i$ is the rate in counter $i$, $f_i(\phi)$ is the value of the 
Fourier series for counter $i$ at angle $\varphi$ and $\sigma_i$ is the 
uncertainty on $r_i$, derived assuming Poisson statistics from the scaler 
value.
The minimization is done numerically
by stepping through a set of 180 $\varphi$ values spaced by 2 degrees.

\begin{figure}[ht]
\centerline{\includegraphics[width=1.2\linewidth]{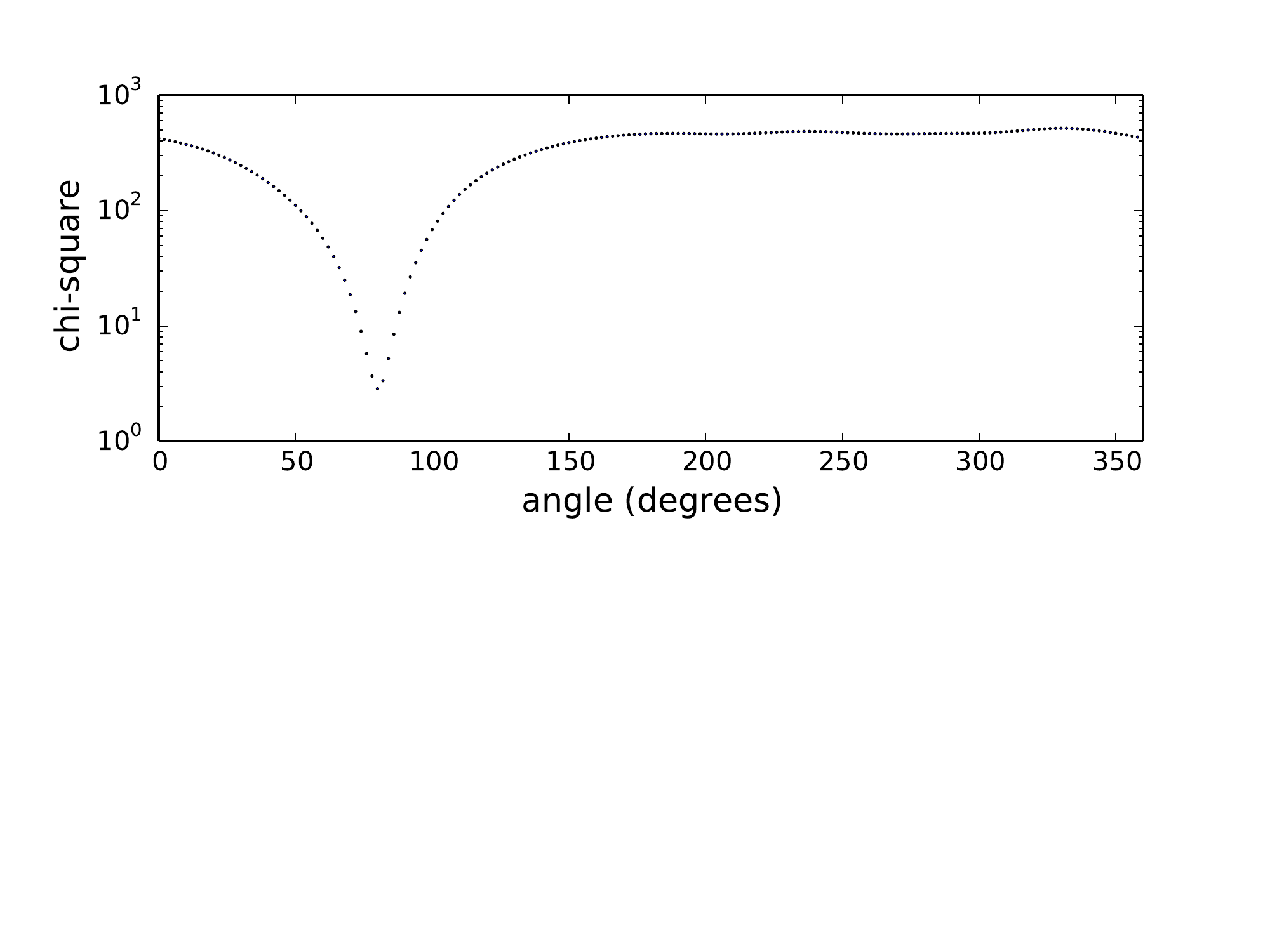}}
\vspace*{-3.0cm}
\caption{$\chi^{2}$/DOF  values as a function of angle.
The minimum occurs at the angle corresponding to the direction of the 
source.}
\label{chi2}
\end{figure}

The technique is illustrated in Figure~\ref{chi2} where the $\chi^{2}$ 
per degree-of-freedom value is
plotted vs. angle for the 180 trials. 
For this curve, 10 two-second data sets were combined to simulate a 20-second
integration time. 
A single minimum is clearly visible. 
This highlights the utility of this method. 
The simple shape of the curve provides powerful visual feedback to the user 
since multiple dips or a single but broader dip would indicate that there are
multiple sources, an extended source, or levels of background comparable to the source intensity. 
The time spent in calculating the 180 trials, using look-up tables for much 
of the work, is negligable in comparison 
to the data acquisition time.

\begin{figure}[ht]
\centerline{\includegraphics[width=1.2\linewidth]{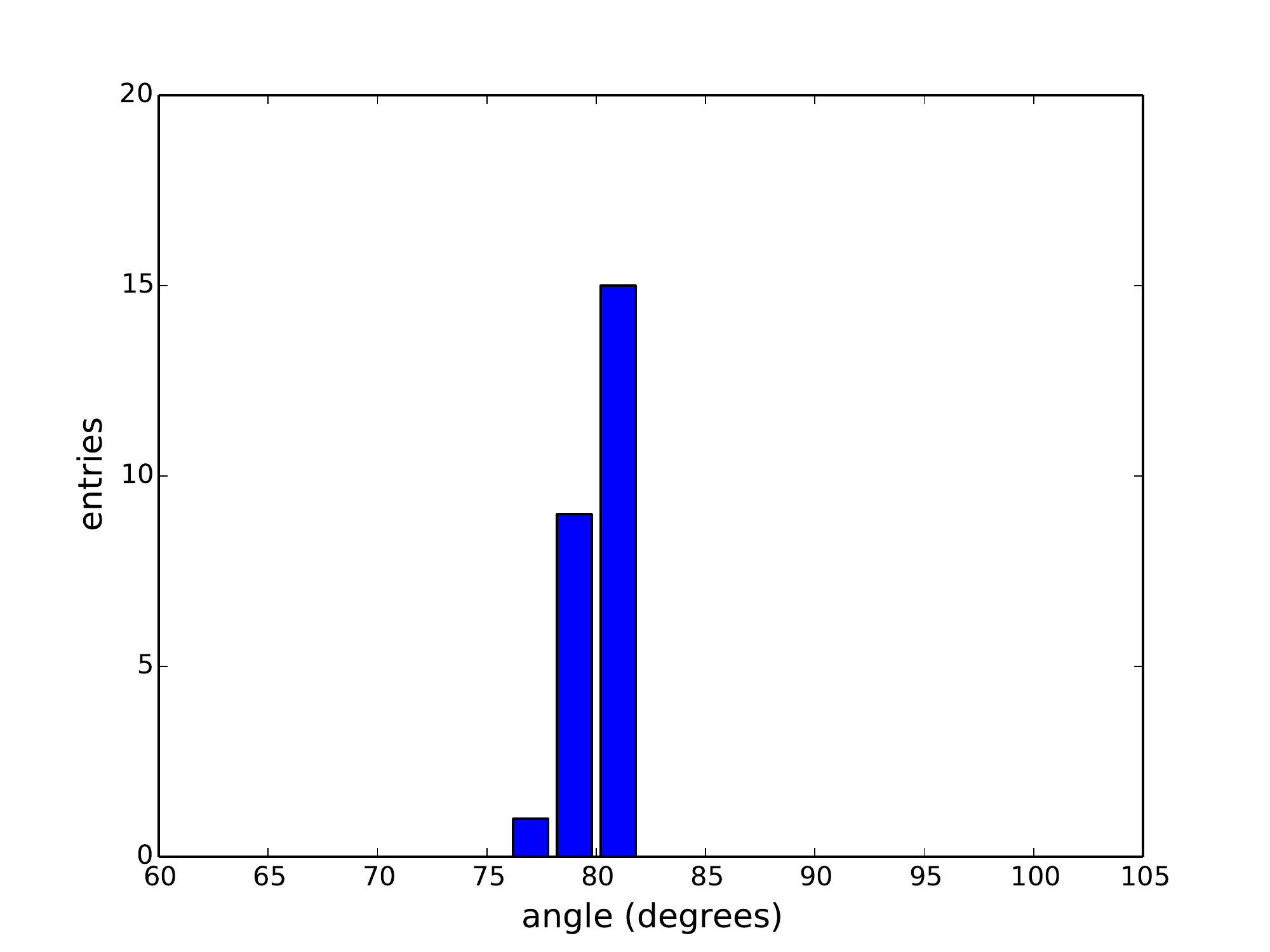}}
\vspace*{0.0cm}
\caption{Distribution of phi values for 25 20-second integrations made at 
at the same angle.}
\label{phi-histo}
\end{figure}

The step value corresponding to the smallest $\chi^{2}$ value is used as the
estimator of the angle.
To obtain 
an estimate of the resolution of this procedure, the process is repeated
24 times using the rest of the data. 
A histogram of the extracted $\varphi$ values, for a single source position, 
is shown in 
Figure~\ref{phi-histo}.
We use the means and RMS values from a series of 16 such distributions, each 
made with the source at a different angle, to 
investigate linearity and resolution.

\begin{figure}[ht]
\centerline{\includegraphics[width=1.2\linewidth]{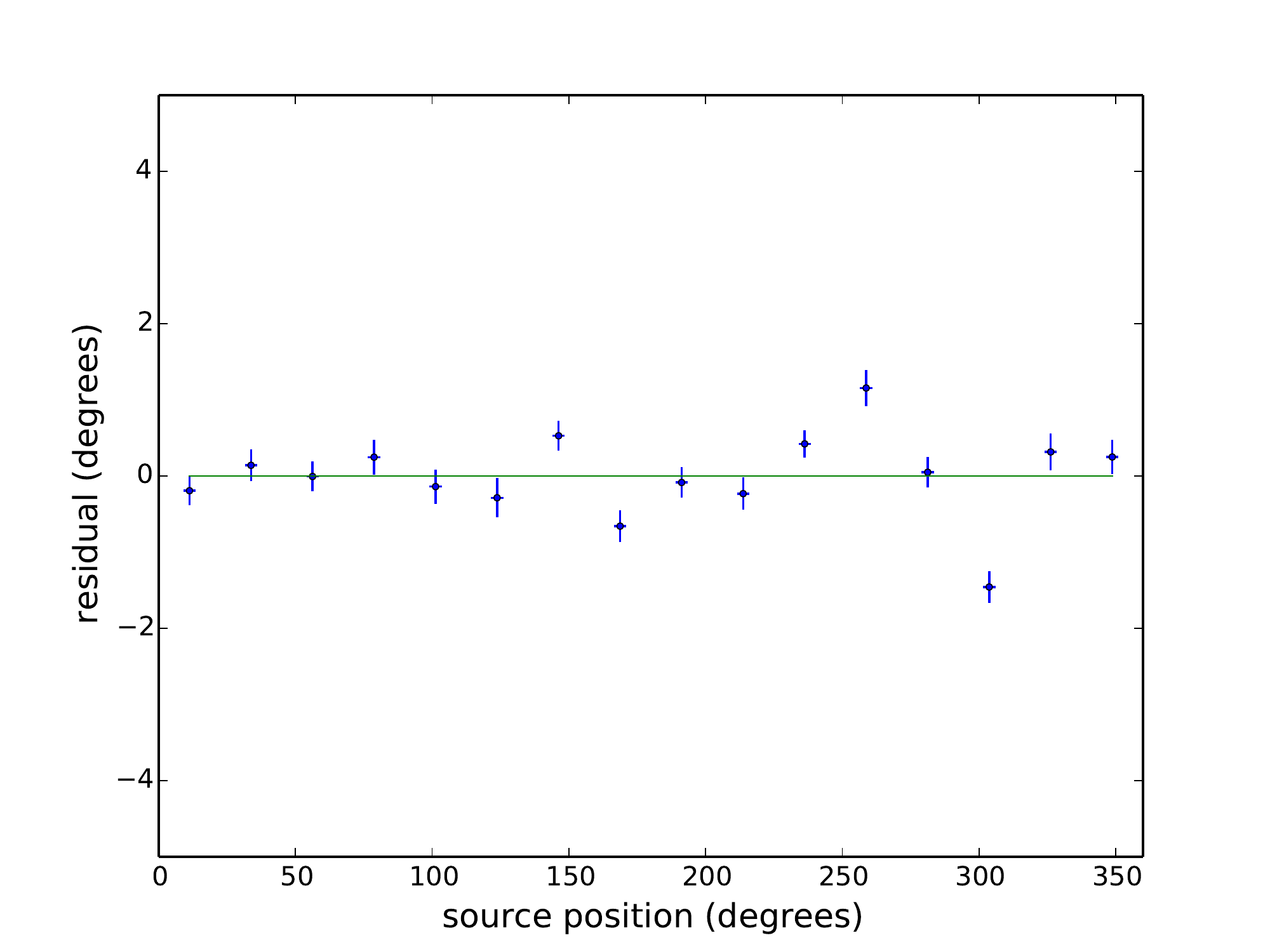}}
\vspace*{0.0cm}
\caption{Residuals from a linear fit of estimated angle vs. 
known source position. The horizontal line is to guide the eye.}
\label{linear}
\end{figure}

A linear function of the angle to the source can be fit to
the means of distributions like the one shown in Figure~\ref{phi-histo}.
Figure~\ref{linear} shows the residuals from such a fit vs. the source
position. 
There is no evidence for 
non-linearity.

\begin{figure}[ht]
\centerline{\includegraphics[width=1.2\linewidth]{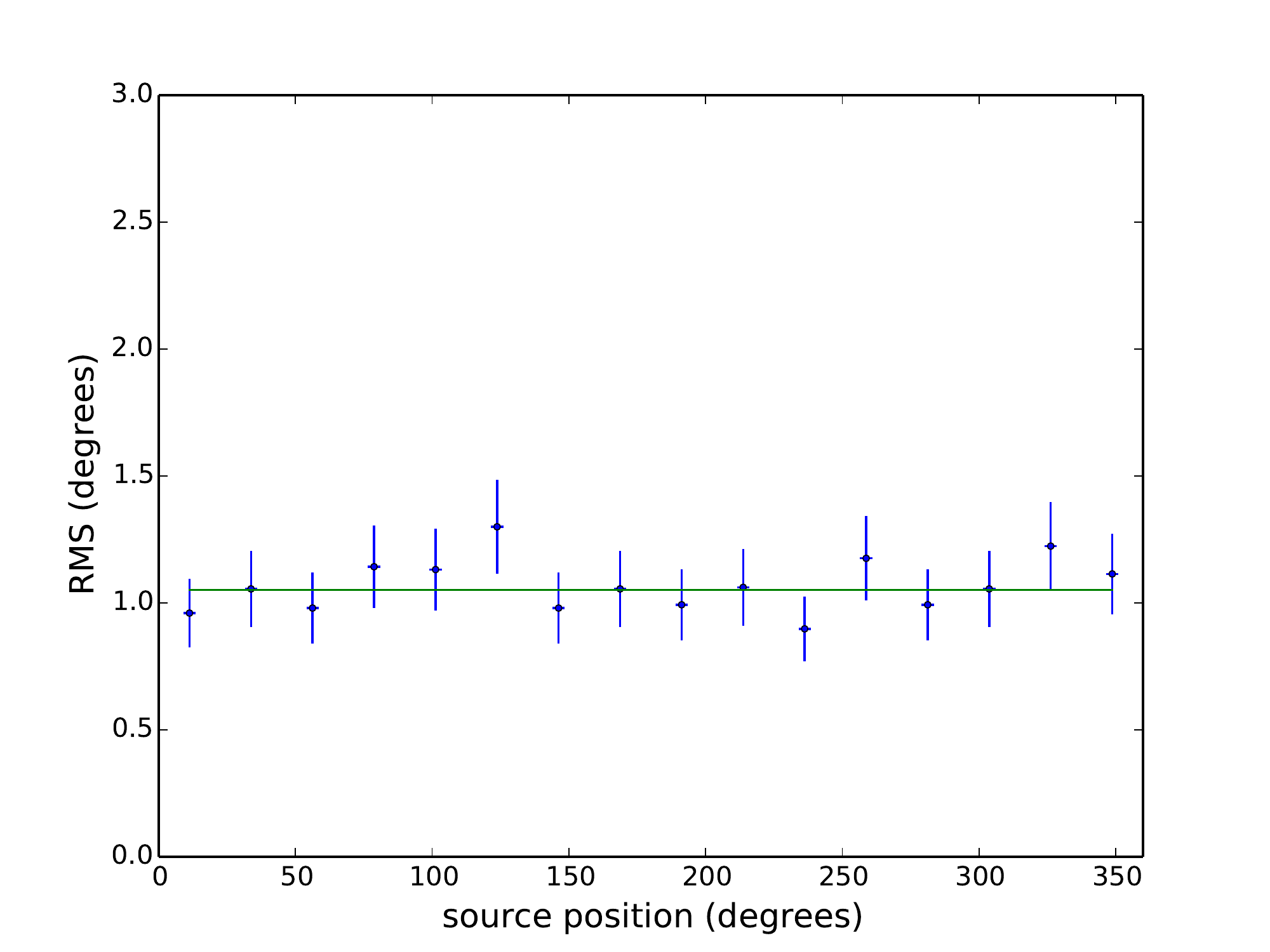}}
\vspace*{0.0cm}
\caption{RMS of the estimated-angle distribution vs. source position.}
\label{RMS}
\end{figure}

Figure~\ref{RMS} shows the RMS values from histograms like the one 
in Figure~\ref{phi-histo} vs. the source position, indicating that 
there is no systematic trend to the angular resolution as a function of 
source angle. 
The mean value is 1.05 $\pm$ 0.02 degrees for our 20-second integration time.

\begin{figure}[ht]
\centerline{\includegraphics[width=1.2\linewidth]{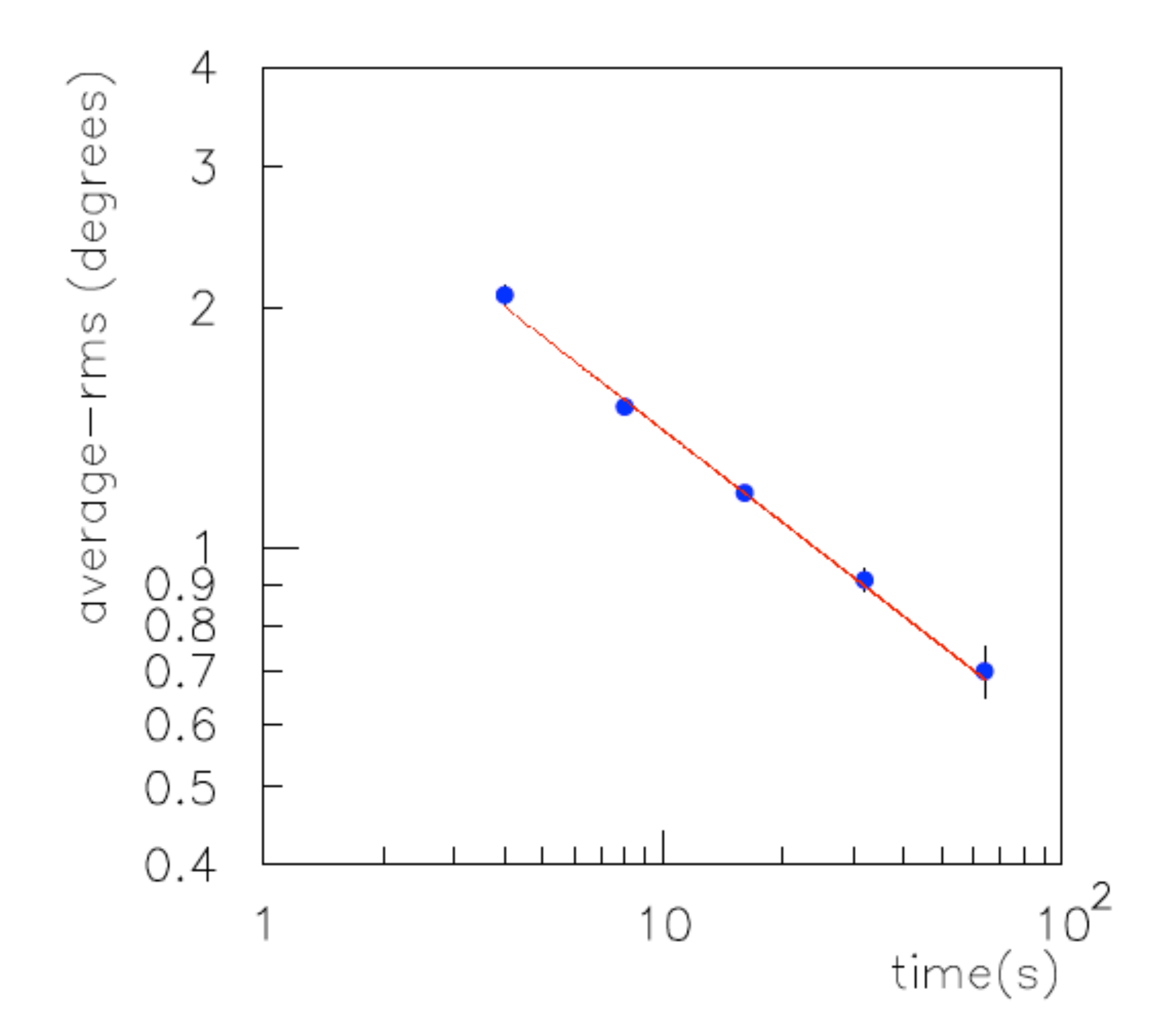}}
\vspace*{0.0cm}
\caption{Average RMS of the estimated-angle distributions as a function of 
the integration time. 
The power-law fit has an index of -0.39 $\pm$ 0.02}
\label{res-plot}
\end{figure}

 To study the effect of integration time on the angular resolution we
                             used a large
               set of simulated data made by generating
random numbers having the same means and widths as the count rates in our
                              data set.
We used these, along with the same calibration parameters that were used
   for obtaining the results in Figures~\ref{linear} and~\ref{RMS},
to perform the same analysis on data sets with different integration times.
         The results are displayed in Figure~\ref{res-plot}.
The angular resolution improves with statistics but at a rate less than the
square-root of the integration time - the power law fit gives an index
                         of -0.39 $\pm$ 0.02.

The curve shows no sign of reaching a constant value as integration time 
increases.
This is because we are plotting the RMS values of distributions like that in
Figure~\ref{phi-histo}. 
Once these drop below a certain level 
the resolution is dominated by the 
systematic scatter like that seen in Figure~\ref{linear}. 
As the error bars on the points in that plot decrease with increasing
statistics one can see the systematics more clearly. 
These can arise from differences in the modules, errors in construction 
geometry and other sources - to attain better than  
degree-scale accuracy requires
more attention to detail and is beyond the scope of this study.

\subsection{Background Considerations}

We have investigated the behaviour of the instrument in the presence of 
background from naturally occuring radioactive material (NORM).
We used the same data as in the previous section but for each set of readings
we added extra counts to the number reported by each of the eight modules.
The amount was simply a Poisson-fluctuated fraction of the eight-module 
average. 
Given that this background is constant, up to statistical fluctuations, 
for each module it adds an overall offset but does not affect the 
modulation effect except to enlarge the statistical uncertainties on the 
count rates.

\section{Multiple Sources}
We investigated the effect of multiple sources at different locations 
with simulations and with 
test runs made with two sources of comparable intensity.  
The effect of extra sources is to distort the $\chi^2$ vs. angle curve, 
like the one shown in Figure~\ref{chi2}, immediately alerting the user that 
there are complications. 

\begin{figure}[ht]
\centerline{\includegraphics[width=1.2\linewidth]{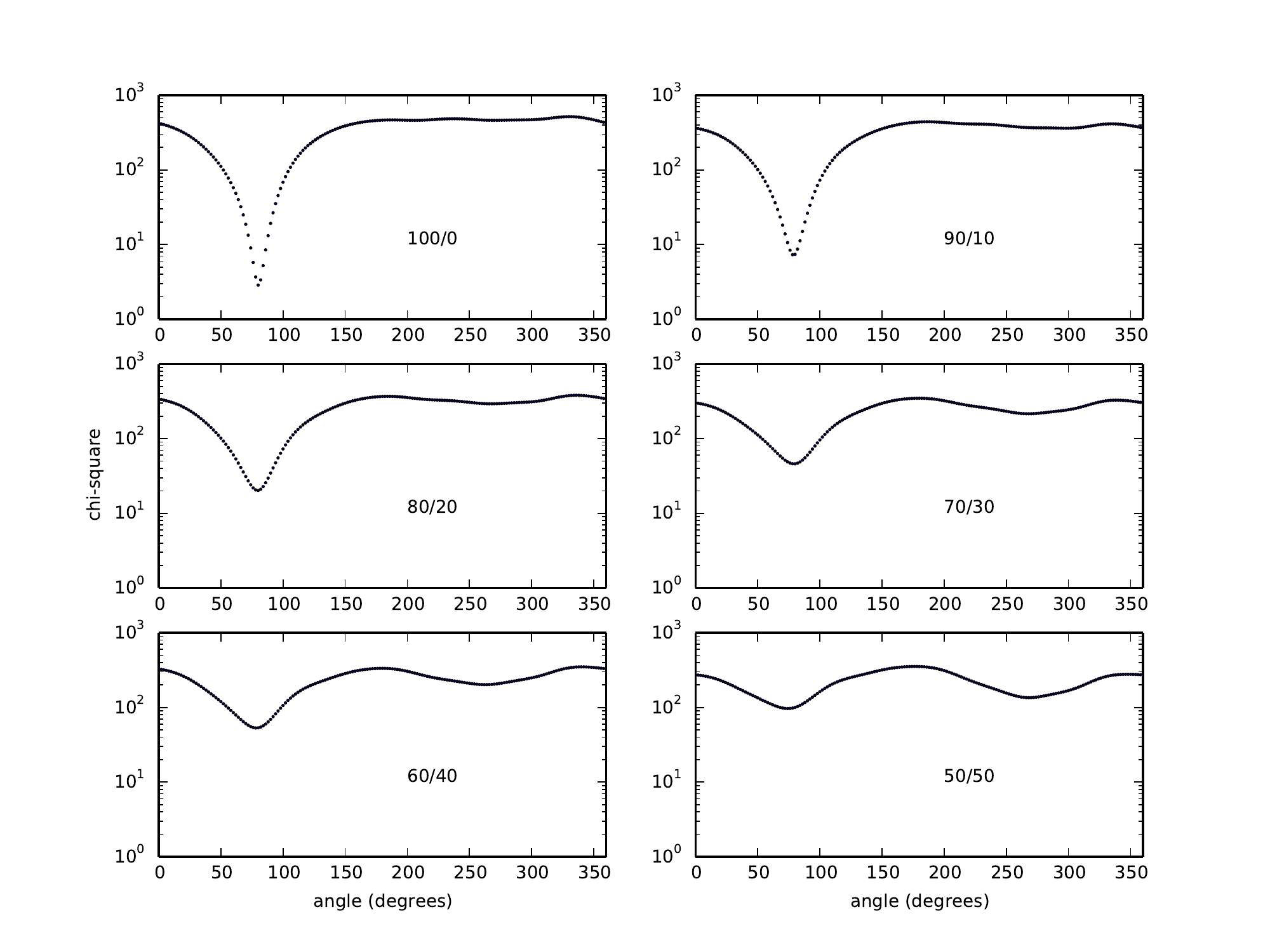}}
\vspace*{0.0cm}
\caption{Simulations of the effect of an additional source located 180 degrees from the main source.
In each panel, the relative intensities of the sources are given by the numbers displayed.
The integration time is fixed at 20 seconds.}
\label{two_source}
\end{figure}

This is illustrated in Figure~\ref{two_source} where we have combined our test files from two 
different angles to simulate the effects of two sources of different strengths.
In the upper left panel the $\chi^2$ vs. angle for a single source is plotted.
In subsequent panels the effects of a source 180 degrees away are included, with the relative 
strength of the second source increasing linearly, until the lower right panel 
where the two sources have the same intensity.

To extract quantitative results a more sophisticated analysis where a hyper-grid involving a number of sources with 
different locations and intensities is searched could be used.  
If the sources are isotopically different, a cut on a spectral feature 
such as the photopeak seen in Figure~\ref{spectrum} could be used to reduce 
the search space.
However, this would require more sophisticated electronics.

\section{Behaviour at Greater Distances}

The motivation for developing this instrument was to aid in the deployment of 
a Compton imager~\cite{laurel,pat,audrey,laurel-2}. 
A design specification for that imager was to locate a 10 mCi point source at
a distance of 40 m to within a degree within one minute.

The studies performed here were done in a small laboratory and necessarily 
involved a geometry where the distance to the source was of the same order 
as the size of the detector. 
For the application envisaged the geometry will be `far-field' so the 
calibration 
curves seen in Figure~\ref{modules-1-8} will be slightly modified; the
modulation features become sharper.

We note that given the attenuation in air of 662 keV photons, and the reduction 
in solid angle, a Cs-137 source 
would need to have an activity of approximately 37 mCi to produce 
the same count rates as those in our tests. 
This means that to determine the direction to a 10 mCi source with 
the resolution that we have achieved in the 
laboratory, we would need integration times that are longer than the 20 
seconds we chose for our studies. 
However the main task of the instrument is to provide a rough estimate 
($\pm$ 5 degrees) of the
direction in which to point the imager so a 20-second integration time is
expected to be sufficient.
A less accurate angle, provided more quickly, is preferable.

\section{Conclusions}

We have developed a simple detector using eight plates of CsI(Tl) 
scintillator to determine
the azimuthal position of a source of gamma rays. 
The detection algorithm relies on the relative count rates for
the different plates, which have a modulation that is well described by 
an 11-component Fourier series.
In the case of a single point source the addition of uniform background 
degrades the angular resolution but does not introduce systematic 
deviations.
If more than one source is present, the relative count rates do not follow
the canonical modulation, but some information on the sources and their 
directions may be recovered depending on their number, locations and 
relative intensities.

The angular resolution of the detector 
improves with
counting statistics and therefore depends on the source strength and
distance and on the level of uniform background. 
Our measurements indicate that for 20-second integration times, a 
resolution of approximately one degree can be obtained for a 
10 $\mu$Ci Cs-137 source at a distance of 80 cm in an environment where the
NORM background is at a level comparable to the source flux.
This meets our design criteria for a device that can aid in deciding 
the optimal deployment direction for a Compton imager.

\section{Acknowledgments}

Undergraduates at McGill University have tested similar devices during 
the development of this instrument~\cite{mcgill_journal}.  
This work has been supported through
funding from the Natural Sciences and Engineering Research Council
(NSERC) the Chemical, Biological, Radiological-Nuclear and Explosives,
Research and Technology Initiative (CRTI Project 07-0193RD).

\end{document}